\documentclass[prd,preprint,
superscriptaddress,showpacs,nofootinbib,%
tightenlines%,twocolumn
]{revtex4}
\usepackage{epsfig}
\usepackage{color}
\usepackage{slashed}
\usepackage{amssymb}
\usepackage{amsmath}

\newcommand{\ben}{\begin{displaymath}}
\newcommand{\een}{\end{displaymath}}
\newcommand{\be}{\begin{equation}}
\newcommand{\ee}{\end{equation}}
\newcommand{\bea}{\begin{eqnarray}}
\newcommand{\eea}{\end{eqnarray}}

\begin{document}

\title{Lambda-nucleon scattering in  baryon chiral perturbation theory}
\author{X.-L.~Ren}
 \affiliation{Ruhr University Bochum, Faculty of Physics and Astronomy,
Institute for Theoretical Physics II, D-44870 Bochum, Germany}
\author{E.~Epelbaum}
 \affiliation{Ruhr University Bochum, Faculty of Physics and Astronomy,
Institute for Theoretical Physics II, D-44870 Bochum, Germany}
\author{J.~Gegelia}
\affiliation{Ruhr University Bochum, Faculty of Physics and Astronomy,
Institute for Theoretical Physics II, D-44870 Bochum, Germany}
 \affiliation{Tbilisi State  University,  0186 Tbilisi,
 Georgia}

\date{15 January, 2020}

\begin{abstract}
We calculate the lambda-nucleon scattering phase shifts and mixing
angles by applying time-ordered
perturbation theory to the manifestly Lorentz-invariant formulation
of SU(3) baryon chiral perturbation theory.  Scattering amplitudes are
obtained by solving the corresponding coupled-channel integral
equations that have a
milder ultraviolet behavior compared to their non-relativistic
analogs.    
This allows us to consider the removed cutoff limit in our leading-order
calculations also in the $^3P_0$ and $^3P_1$ partial waves. We find
that, in the framework we are using, at least some part of the
higher-order contributions to the baryon-baryon potential in these
channels needs to be treated nonperturbatively and demonstrate how
this can be achieved  in a way consistent with quantum field
theoretical renormalization for the leading contact interactions. 
We compare our results with the ones of the non-relativistic approach
and lattice QCD phase shifts obtained for non-physical pion masses.

\end{abstract}

\maketitle

\section{\label{introduction}Introduction}

Nuclear systems with non-vanishing strangeness play an important role in the study of nuclear, particle and astrophysics.
Hyperon-nucleon
(YN) interactions are crucial for understanding hypernuclear binding.
Experiments investigating the YN,
hyperon-hyperon (YY) and cascade-nucleon ($\Xi$N) interactions 
are carried out at various laboratories such as CERN,
DA$\Phi$NE, GSI, JLab, J-PARC, KEK, MAMI, RHIC and will also be
performed at the future FAIR facility. For reviews on the subject of hypernuclear physics see
Refs.~\cite{Pochodzalla:2011rz,Esser:2013aya,Feliciello:2015dua,Gal:2016boi}. 

Lattice QCD is another  valuable source of information on YN  and YY
interactions \cite{Miyamoto:2016hqo,Doi:2017zov,Nemura:2017vjc,Sasaki:2018mzh,Beane:2013br,Beane:2012ey,Beane:2012vq,Beane:2010em,Beane:2006gf,Beane:2008dv,Yamazaki:2012hi,Yamazaki:2015asa,Hanlon:2018yfv}. 
While some lattice simulations are approaching the physical values of the light quark masses,  most of the available lattice QCD calculations 
still correspond to unphysically large values and, therefore, require extrapolations to their physical values.

Chiral effective field theory (ChEFT) is a
natural framework for analyzing low-energy properties of (hyper)nuclei and
performing  chiral extrapolations. Chiral perturbation theory for systems involving two and more nucleons has been initiated in Refs.~\cite{Weinberg:rz,Weinberg:um}.
In that formulation,
%ChEFT approach to few-baryon systems
%the
power counting rules
are applied to the effective potentials, and the scattering amplitude  is then obtained by solving the
Lippmann-Schwinger (LS) or Schr\"odinger equations.
Reviews of ChEFT in the few-body sector can be found in
Refs.~\cite{Bedaque:2002mn,Epelbaum:2005pn,Epelbaum:2008ga,Machleidt:2011zz,Epelbaum:2012vx}.

Chiral EFT
for baryon-baryon (BB) interactions in the strange sector has been formulated in Refs.~\cite{Meissner:2016ood,Haidenbauer:2015zqb,Petschauer:2015nea,Haidenbauer:2014rna,Haidenbauer:2013oca,Haidenbauer:2011za,Haidenbauer:2011ah,
Haidenbauer:2009qn,Polinder:2007mp,Haidenbauer:2007ra,Polinder:2006zh,Haidenbauer:2018gvg,Haidenbauer:2014uua,Haidenbauer:2017sws}
using the non-relativistic framework.
In these studies,  the standard Weinberg
power counting for nucleon-nucleon (NN) interactions is extended to the
non-zero strangeness sectors of the baryon-baryon interaction. 
Ultraviolet divergences of the LS equation have been taken care of by applying
finite-cutoff regularitazion  using 
exponential cutoffs in the range of $500\ldots
700$~MeV.

For nucleon-nucleon scattering, a modified Weinberg  approach with an improved ultraviolet behavior has been proposed in
Ref.~\cite{Epelbaum:2012ua}. This novel framework uses time-ordered
perturbation theory (TOPT) applied to the manifestly Lorentz-invariant
effective Lagrangian and leads to the Kadyshevsky equation for the scattering amplitude
\cite{kadyshevsky}.
It has been explored in the non-strange sector \cite{Epelbaum:2013ij,Epelbaum:2013naa,Epelbaum:2015sha}, 
also with a different treatment of Dirac spinors and using an alternative power counting \cite{Ren:2016jna}.
First applications of a similar formalism to baryon-baryon systems with non-zero strangeness
can be found in
Refs.~\cite{Li:2016mln,Ren:2018xxd,Li:2016paq,Li:2018tbt,Song:2018qqm}.\footnote{For an application of the  modified Weinberg  approach to hadronic
  molecules see Ref.~\cite{Baru:2015tfa}.}

Recently, we have worked out in details the  diagrammatic rules of TOPT
for particles with non-zero spin and for interactions involving time
derivatives \cite{Baru:2019ndr}.  In this paper, we apply the
resulting framework to 
the strangeness $S=-1$ sector of baryon-baryon scattering and focus,
in particular, on lambda-nucleon and sigma-nucleon scatterings.

Our paper is organized as follows: in section~\ref{effective_Lagrangian} we specify the effective Lagrangian required for our calculations. 
In section~\ref{intequation}, we consider the system of integral
equations for baryon-baryon scattering and present the
lambda-nucleon and sigma-nucleon scattering phase
shifts.
Next, in section~\ref{rensubtract}, we discuss the renormalization of
the scattering amplitudes with the next-to-leading order (NLO) contact
interaction potentials treated nonperturbatively. 
The results of our work are
summarized in  section~\ref{conclusions}.

\section{Effective Lagrangian}
\label{effective_Lagrangian}

The starting point of our analysis is 
the manifestly Lorentz-invariant effective Lagrangian of baryon chiral perturbation theory consisting of the purely mesonic, single-baryon and two-baryon parts,
\begin{equation}
{\cal L}_{\rm eff}={\cal L}_{\rm \phi}+{\cal L}_{{\rm \phi B}}+{\cal
L}_{\rm BB}\,. \label{inlagr}
\end{equation}
From the purely mesonic sector, we only need the lowest-order Lagrangian 
\cite{Gasser:1984yg} 
\begin{equation}\label{piLag}
{\cal L}^{(2)}_{\rm \phi}=\frac{F_0^2}{4}\mbox{Tr}\left\{ u_\mu
u^\mu +\chi_+\right\} ,
\end{equation}
where
\begin{equation}
 u_\mu =iu^{\dagger} \partial_\mu U u^{\dagger},\quad u^2=U=\exp\left(\sqrt{2} i \phi/ F_0 \right) ,\quad
\chi_{\pm}= u^\dagger\chi u^\dagger
\pm u \chi^\dagger u, \quad \chi = 2 B_0 s ,
\label{defB0}
\end{equation}
 and $\phi$ is the irreducible octet representation of ${\rm SU}(3)_f$ for the Goldstone bosons,
\begin{equation}
 \phi  = \left(%
\begin{array}{ccc}
\frac{ \pi^0}{\sqrt{2}}+\frac{\eta }{\sqrt{6}}  & \pi^+ & K^+\\
\pi^- & \frac{-\pi^0}{\sqrt{2}}+\frac{\eta }{\sqrt{6}}  & K^0\\
K^- & \bar K^0 & -\frac{2 \eta }{\sqrt{6}}\\
\end{array}%
\right).
\label{defP}
\end{equation}
Here $F_0$ stands for the meson decay constant in the chiral
limit while $s$ is the external scalar source that gives rise
to the quark (and Goldstone-boson)
masses. For the purposes of the current work we
  switch off  all other external sources and
consider the isospin-symmetric case of $m_u=m_d\neq {m_s}$. The constant $B_0$ in Eq.~(\ref{defB0}) is related to the quark condensate.

The leading-order (LO) Lagrangian of the single-baryon sector is given by
\begin{equation}
{\cal L}_{{\rm \phi B}}^{(1)}=\mbox{Tr} \left\{ \bar {\rm B} \left( i\gamma_\mu D^\mu -m \right)  {\rm B} \right\}  +\frac{D/F}{2}
\mbox{Tr} \left\{\bar {\rm B} \gamma_\mu \gamma_5 [u^\mu,{\rm B}]_{\pm}\right\}  ,
\label{lolagr}
\end{equation}
where ${\rm B}$ is the irreducible
octet representation of ${\rm SU}(3)_f$ involving baryon fields, 
\begin{equation}
 {\rm B} = \left(%
\begin{array}{ccc}
\frac{ \Sigma^0}{\sqrt{2}}+\frac{\Lambda }{\sqrt{6}}  & \Sigma^+ & p\\
\Sigma^- & \frac{-\Sigma^0}{\sqrt{2}}+\frac{\Lambda }{\sqrt{6}}  & n\\
\Xi^- & \Xi^0 & -\frac{2 \Lambda }{\sqrt{6}}\\
\end{array}%
\right),
\label{defBbarB}
\end{equation}
$D$ and $F$ are coupling constants corresponding to $[\ldots ]_+$ and
$[\ldots ]_{-}$, respectively,
and $D_\mu{\rm B} = \partial_\mu {\rm B}+[[u^\dagger,\partial_\mu u], {\rm B}] $
is the covariant derivative.
%We need the following two terms from the NLO meson-baryon Lagrangian
%\begin{equation}
%{\cal L}_{{\rm \phi B}}^{(2)}=b_{D/F} \,\mbox{Tr} \left\{ \bar {\rm B} \left[ \chi_+,  {\rm B}\right]_{\pm} \right\}  + b_0 \,
%\mbox{Tr} \left\{ \bar {\rm B} {\rm B}\right\}  \mbox{Tr} \left\{ \chi_+\right\} ,
%\label{nlolagr}
%\end{equation}
%where $b_{D/F}$ and $b_0$ are coupling constants.

The effective baryon-baryon Lagrangian contributing to the
LO BB potential consists of the following terms
\cite{Polinder:2007mp} 
\begin{eqnarray}
{\cal L}_{\rm BB} &=& C^1_i \, \mbox{Tr} \left\{ \bar {\rm B}_\alpha \bar {\rm B}_\beta \left(  \Gamma_i {\rm B}\right)_\beta \left(  \Gamma_i {\rm B}\right)_\alpha  \right\}
+C^2_i \, \mbox{Tr} \left\{ \bar {\rm B}_\alpha  \left(  \Gamma_i {\rm B}\right)_\alpha \bar {\rm B}_\beta \left(  \Gamma_i {\rm B}\right)_\beta  \right\}  \nonumber\\
&+& C^3_i \, \mbox{Tr} \left\{ \bar {\rm B}_\alpha  \left(  \Gamma_i {\rm B}\right)_\alpha  \right\} \mbox{Tr} \left\{ \bar {\rm B}_\beta \left(  \Gamma_i {\rm B}\right)_\beta  \right\}  ,
\label{NNLagrdreg}
\end{eqnarray}
where $C_i^1$, $C_i^2$ and  $C_i^3$ are the coupling constants,  $\alpha$ and $\beta$ are the Dirac spinor indices and
\begin{equation}
\Gamma_1=1, \ \ \  \Gamma_2=\gamma^\mu, \ \ \  \Gamma_3=\sigma^{\mu\nu}, \ \ \  \Gamma_4=\gamma^\mu\gamma_5 .
\label{structures}
\end{equation}
Notice that the $ \Gamma_5 =\gamma_5 $-term starts contributing at NLO.

%In our calculations we use the Dirac spinors with the four-momentum $q$ %and the polarization $\lambda$ 
%\be\label{nuclspin}
% u(q) = \sqrt{\frac{\omega(q,m_N)+m_N}{2m_N}}    \left(\begin{array}{c} \chi \\
%                        \frac{\vec{\sigma}\cdot\vec{q} \ \chi }{\omega(q,m_N)+m_N}
%                 \end{array}\right), 
%\ee 
%where $\omega(q,M):=\left(\vec q\,{  }^2+M^2\right)^{1/2}$ is the particle energy and $\chi$ - the two-component spinor. 
%

\section{ Integral equations for  baryon-baryon scattering }
\label{intequation}

The off-shell baryon-baryon scattering amplitude $T$ satisfies
the integral equation, which can symbolically be written as \cite{Baru:2019ndr}:
\begin{equation}
T=V+ V G\,T\,,
\label{Teq1}
\end{equation}
where $V$ is the effective potential and $G$ is the two-baryon
Green function.
To obtain the scattering amplitudes of processes with strangeness $S=-1$ in the
isospin limit, Eq.~(\ref{Teq1}) is understood as a $2\times 2$ matrix equation,
%(a generalization of the Kadyshevsky equation \cite{kadyshevsky})  
where
\begin{eqnarray}
 T =
 \left(%
\begin{array}{cc}
 T_{\Lambda N,\Lambda N},  & T_{\Lambda N,\Sigma N} \\
 T_{\Sigma N,\Lambda N},   & T_{\Sigma N,\Sigma N} \\
\end{array}%
\right),
\ \ 
V =  \left(%
\begin{array}{cc}
 V_{\Lambda N,\Lambda N},  & V_{\Lambda N,\Sigma N}  \\
 V_{\Sigma N,\Lambda N},   & V_{\Sigma N,\Sigma N} \\
\end{array}%
\right), \ \ 
G =  \left(%
\begin{array}{cc}
G^{\Lambda N}  & 0 \\
0   & G^{\Sigma N}  \\
\end{array}%
\right),
\label{defMatrixEq}
\end{eqnarray}
and the two-body Green functions read 
\begin{equation}
G^{IJ}(E)= \frac{1}{\omega(p_I, m_I) \omega(p_J, m_J)}\, \frac{m_I m_J}{E - \omega(p_I, m_I)-\omega(p_J, m_J)+i \epsilon} \,,
\label{Gij}
\end{equation}
with $m_I$ and $\omega (p_I, m_I) = \left(\vec
  p_I\,{}^2+m_I^2\right)^{1/2}$    
being the mass and  the energy of
the $I$-th baryon with
four-momentum $p_I$.
%
%
%We calculate the baryon-baryon scattering amplitude in
%the center-of-mass system (CMS).  
%We denote 
%the three-momenta of the incoming and
%outgoing baryons  in the CMS by  $\vec p$ and  $\vec p\,'$, respectively.
In the partial wave basis, Eq.~(\ref{Teq1}) leads to
the following coupled-channel equations with the partial wave projected potentials
$V^{IJ,KL}_{l'l,s's,j}\left(E; p',p\right)$,
\begin{eqnarray}
T^{IJ,KL}_{l'l,s's,j}\left(E; {p'}
,p\right) &=& V^{IJ,KL}_{l'l,s's,j}\left(E;p',p\right) +  \nonumber\\
&& \sum_{l'',s'',P,Q}
\int_0^\infty
\frac{d k\,k^2}{2 \pi^2} V^{IJ,PQ}_{l'l'',s's'', j}\left(E;p',k\right) G^{PQ}(E) \, T^{PQ,KL}_{l''l,s''s,j}
\left(E; k ,p\right),
\label{PWEHDR}
\end{eqnarray}
where $IJ, KL$ and $PQ$ label the initial, final and intermediate
particles, respectively,  and
$p \equiv | \vec p \,|$, $p ' \equiv | \vec p\,' |$ and $k \equiv | \vec k
\,|$.  
The indices $l$, $l''$, $l'$ and $s$, $s''$, $s'$ stand for their orbital
angular momentum and spin, respectively, while $j$ denotes the total angular
momentum of the BB states.
Compared to the corresponding Lippmann-Schwinger equation for the same potential, Eq.~(\ref{PWEHDR}) with the Green functions of Eq.~(\ref{Gij})
has a milder UV behaviour.
Therefore, its solutions show less sensitivity to the variation of the
cutoff parameter \cite{Baru:2019ndr}.

We organize the BB potential by applying the standard Weinberg power counting to two-baryon irreducible TOPT diagrams. 
The LO potential consists of the short-range contact interaction part
$V_{{\rm LO},C}^{IJ,KL}$ and the one-meson exchange (OME) contribution \cite{Baru:2019ndr}
%For the  one-meson-exchange contribution to the LO potential we have
 \begin{eqnarray}\label{OBEpotLO}
V_{{\rm LO},M_P}^{IJ,KL} &=& -\frac{f_{IKP} f_{JLP} \,{\cal I}_{IJ,KL} }{2\, \omega(q,M_P) }\,
%\nonumber\\ &\times&
\left[ \frac{ 1}{\omega(q,M_P) +\omega(p_K,m_K) +\omega(p_J,m_J) -E-i\,\epsilon } \right. \nonumber\\
&&{}+ \left. \frac{1}{\omega(q,M_P) +\omega(p_L,m_L) +\omega(p_I,m_I) -E-i\,\epsilon } \right]  \frac{\left(m_I+m_K\right) \left(m_J+m_L\right)}{
   \sqrt{m_I m_J m_K m_L}}
\\
&&{}\times\frac{
   \left(m_K \vec\sigma_1 \cdot\vec p_I - m_I \vec\sigma_1 \cdot\vec p_K
  \right){}  \left( m_L \vec\sigma_2 \cdot\vec p_J  - m_J\vec\sigma_2 \cdot\vec p_L \right)
   }{
   \sqrt{\omega
   \left(p_I,m_I\right)+m_I}\,
   \sqrt{\omega \left(p_J,m_J\right)+m_J}\, \sqrt{\omega
   \left(p_K,m_K\right)+m_K}\, \sqrt{\omega \left(p_L,m_L\right)+m_L}},  \nonumber
\end{eqnarray}
where $q=p_I-p_K =p_L-p_J$ is the four-momentum transfer. 
  The isospin factors ${\cal I}_{IJ,KL} $ and the values of $f_{IKP}$ can be found in Refs.~\cite{Haidenbauer:2011ah,Haidenbauer:2007ra}.

It is straightforward to obtain from the Lagrangian of Eq.~(\ref{NNLagrdreg}) the expressions for the contact interactions, which we include at LO according to Ref.~\cite{Baru:2019ndr}.  
They are identical to those of the non-relativistic approach and can be found in Refs.~\cite{Polinder:2006zh,Polinder:2007mp,Ren:2016jna,Li:2016mln,Li:2018tbt}. 

Due to the SU(3) flavour symmetry assumed in Ref.~\cite{Polinder:2006zh}, there are two LECs in the contact terms of $^1S_0$ and $^3S_1$ partial waves for the $\Lambda N$-$\Sigma N$ coupled channels, respectively. 
However, as argued in Ref.~\cite{Haidenbauer:2019boi}, 
the SU(3) symmetry is always broken by the one-meson-exchange
potential in calculations using the actual meson and baryon
masses. Therefore, we have to introduce an extra contact term in each
of the $^1S_0$ and $^3S_1$ partial waves in order to be able to carry out renormalization.   
Specifically, we consider the contact interactions of the following form
\begin{equation}
	V_{\mathrm{LO},C}^{^1S_0} = \begin{pmatrix}
		C^{^1S_0}_{\Lambda N,\,\Lambda N} & C^{^1S_0}_{\Lambda N,\,\Sigma N}  \\
C^{^1S_0}_{\Lambda N,\, \Sigma N}  & C^{^1S_0}_{\Sigma N,\,\Sigma N}  
	\end{pmatrix}, \quad
	V_{\mathrm{LO},C}^{^3S_1} = \begin{pmatrix}
		C^{^3S_1}_{\Lambda N,\,\Lambda N} & C^{^3S_1}_{\Lambda N,\,\Sigma N}  \\
C^{^3S_1}_{\Lambda N,\, \Sigma N}  & C^{^3S_1}_{\Sigma N,\,\Sigma N}  
	\end{pmatrix}.
\end{equation}
\noindent
To determine these LECs we fit them to the low-energy phase shifts
quoted in Ref.~\cite{Haidenbauer:2019boi}, where non-relativistic
chiral EFT potentials up to NLO were employed to describe the scattering
observables. The values of LECs are listed in Table~\ref{tab:LECs}.

\begin{table}[b]

  \caption{The fitted LECs of the $^1$S$_0$ and $^3$S$_1$ partial waves in the $\Lambda N$-$\Sigma N$ scattering. The momentum cutoff is taken as $\Lambda=20$ GeV.}
  \label{tab:LECs}
 \begin{tabular*}{\textwidth}{@{\extracolsep{\fill}}cccc} 
  \hline\hline 
    $C^{^1S_0}_{\Lambda N,\Lambda N} [10^4 \mathrm{GeV}^{-2}]$ & -0.0186  & $C^{^3S_1}_{\Lambda N,\Lambda N} [10^4 \mathrm{GeV}^{-2}]$ & -0.0124 \\
    $C^{^1S_0}_{\Lambda N,\Sigma N} [10^4 \mathrm{GeV}^{-2}]$ & 0.00361 &  $C^{^3S_1}_{\Lambda N,\Sigma N} [10^4 \mathrm{GeV}^{-2}]$ & -0.00499\\
    $C^{^1S_0}_{\Sigma N,\Sigma N} [10^4 \mathrm{GeV}^{-2}]$ & -0.00849 & $C^{^3S_1}_{\Sigma N,\Sigma N} [10^4 \mathrm{GeV}^{-2}]$ & -0.0129 \\
    \hline\hline   
  \end{tabular*}
\end{table}

\begin{figure}[tb]
\includegraphics[width=\textwidth]{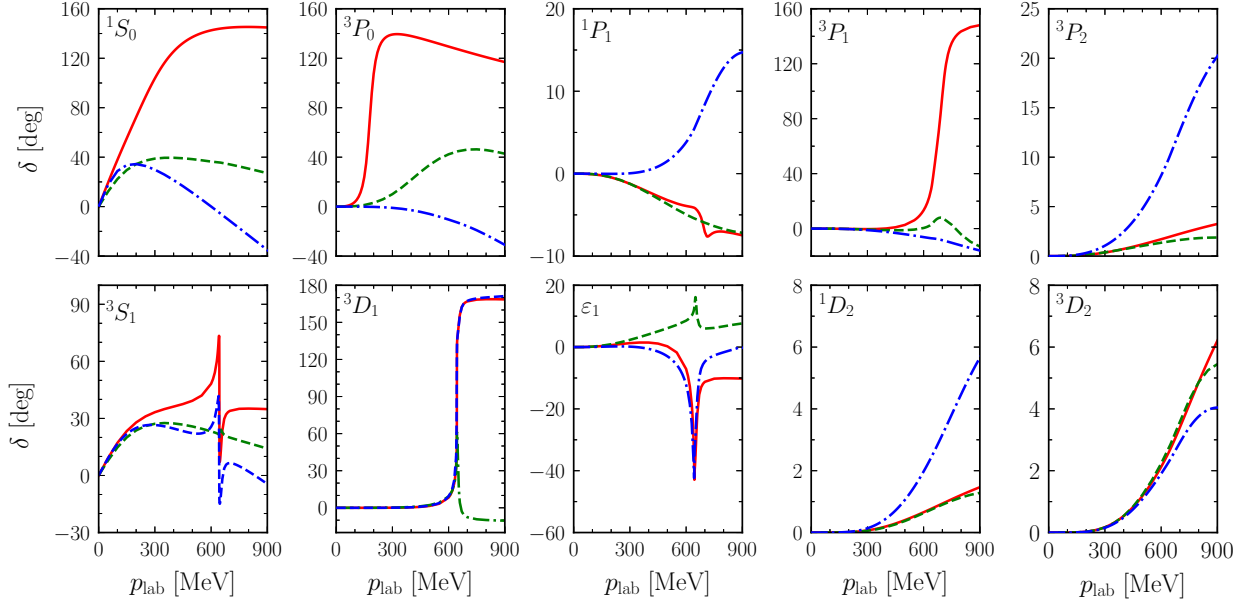}
  \caption{Phase shifts of $\Lambda N$ scattering. Red lines represent our LO results with the cutoff $\Lambda=20$~GeV, green dashed lines and blue dot-dashed lines 
  denote the non-relativistic results (obtained with $\Lambda=0.6$ GeV) at LO~\cite{Polinder:2006zh} 
  and NLO~\cite{Haidenbauer:2019boi}, respectively.}
  \label{LNps:fig}
\end{figure}

\begin{figure}[tb]
\includegraphics[width=\textwidth]{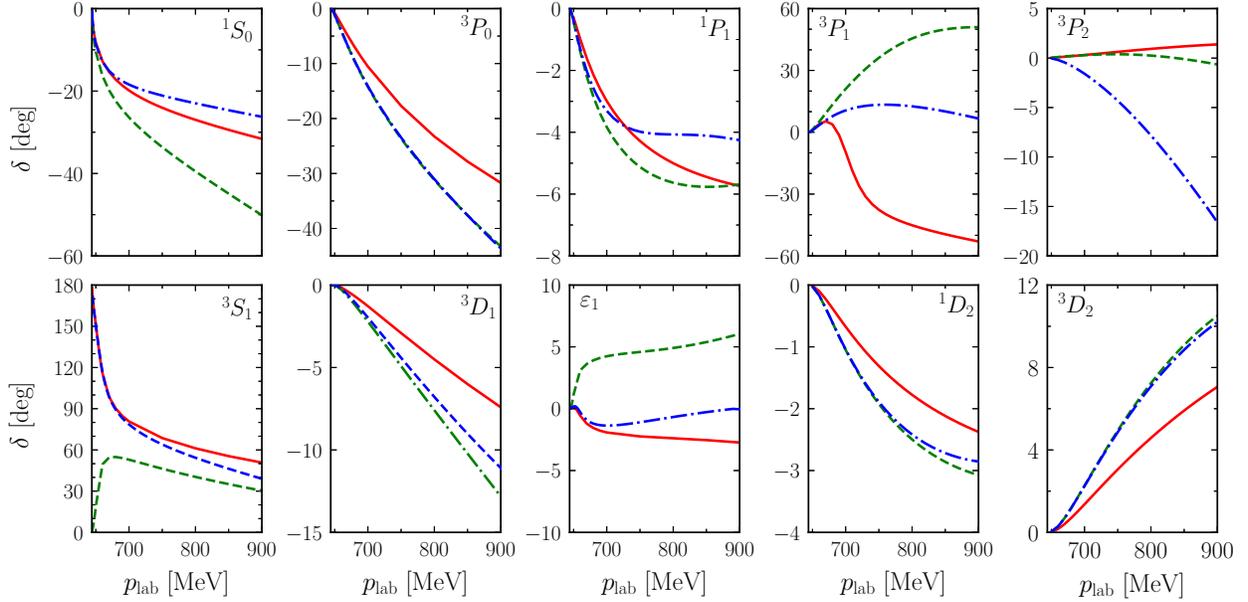}
  \caption{Phase shifts of $\Sigma N$ scattering. Same description of curves as in Fig.~\ref{LNps:fig}.}
  \label{SNps:fig}
\end{figure}

Our results for the baryon-baryon scattering phase shifts in
strangeness $S=-1$ sector with isospin $1/2$ are plotted in
Figs.~\ref{LNps:fig} and \ref{SNps:fig}.\footnote{In all
  figures in this paper, $p_{\rm lab}$ is defined as the momentum of
  the incoming $\Lambda$-particle in the laboratory system with
  the nucleon at rest.} 
The obtained phase shifts for
partial waves without contact interactions at  LO are similar to those
of the non-relativistic approach except for the $^3P_0$ partial wave
in the
$\Lambda N$ channel and the $^3P_1$ partial wave in the $\Lambda N$
and $\Sigma N$ channels, which strongly deviate from the 
phase shifts of Ref.~\cite{Haidenbauer:2019boi}.   
The phase shifts and mixing angles for the $^3$S$_1$-$^3$D$_1$ coupled channels are consistent with the NLO 
results of Ref.~\cite{Haidenbauer:2019boi}.
%Besides, for $^1$S$_0$ and $^3$S$_1$-$^3$D$_1$ partial waves our results of OMEP are similar to those of the non-relativistic approach, while the phase shifts of the $^3P_0$ partial wave of $\Lambda N$ scattering and the $^3P_1$ partial wave of the $\Lambda N$ and $\Sigma N$ scatterings show different behaviour.
The observed large differences between our LO results and the  phase
shifts in the  $^1$S$_0$, $^3$P$_0$ and  $^3$P$_1$ channels as well as the
large NLO corrections in these partial waves found in
Ref.~\cite{Haidenbauer:2019boi} suggest that certain contributions to  
the potential beyond LO in these channels need to be treated
nonperturbatively. In the next section, we will show how the
contributions of the P-wave contact interactions can be resumed
nonperturbatively in the way compatible with EFT. As for the $^1S_0$
partial wave, a more involved treatment similar to that of
Ref.~\cite{Epelbaum:2015sha} in the nonstrange sector is needed. This 
case will be treated in a separate publication.

We also calculated the scattering lengths and the effective range parameters of
the $\Lambda N$ scattering for the $^1$S$_0$ and $^3$S$_1$ partial
waves. Our results, given in Table~\ref{tab:ERE}, agree reasonably
well with those of the non-relativistic approach except the effective
range of the $^1S_0$ partial wave, which has different sign, as
already visible from the corresponding phase shifts shown in
Fig.~\ref{LNps:fig}. 
 
\begin{table}[bt]
  \caption{The scattering lengths and effective ranges in the $\Lambda N$ scattering for the $^1$S$_0$ and $^3$S$_1$ partial waves. The non-relativistic results from LO and NLO studies are also listed.}
  \label{tab:ERE}
 \begin{tabular*}{\textwidth}{@{\extracolsep{\fill}}cccc} 
  \hline\hline 
     & Rel.-LO, $\Lambda=20$~GeV & NR-LO~\cite{Polinder:2006zh},$\Lambda=0.6$~GeV & NR-NLO~\cite{Haidenbauer:2019boi}, $\Lambda=0.6$~GeV\\
    \hline
    $a_{^1S_0} [{\rm fm}]$ & -2.94  & -1.91 & -2.91 \\
    $r_{^1S_0} [{\rm fm}] $ & -1.44 &  1.40 & 2.78\\
    \hline 
    $a_{^3S_1} [{\rm fm}] $ & -1.41 & -1.23 & -1.41 \\
    $r_{^3S_1} [{\rm fm}] $ & 1.61 &  2.13 & 2.53 \\
    \hline\hline   
  \end{tabular*}
\end{table}

\begin{figure}[bt]
\includegraphics[width=0.32\textwidth]{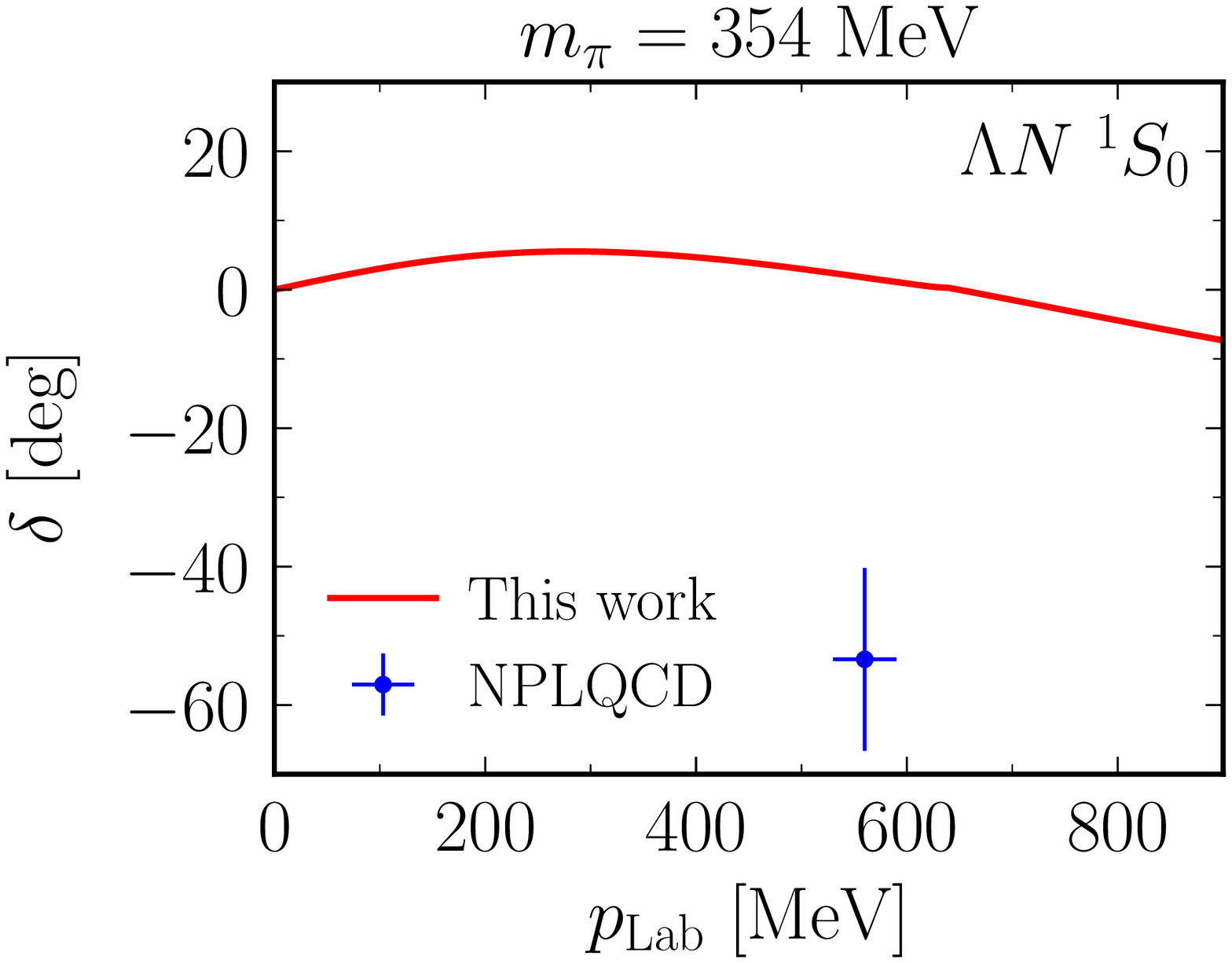}~~
\includegraphics[width=0.32\textwidth]{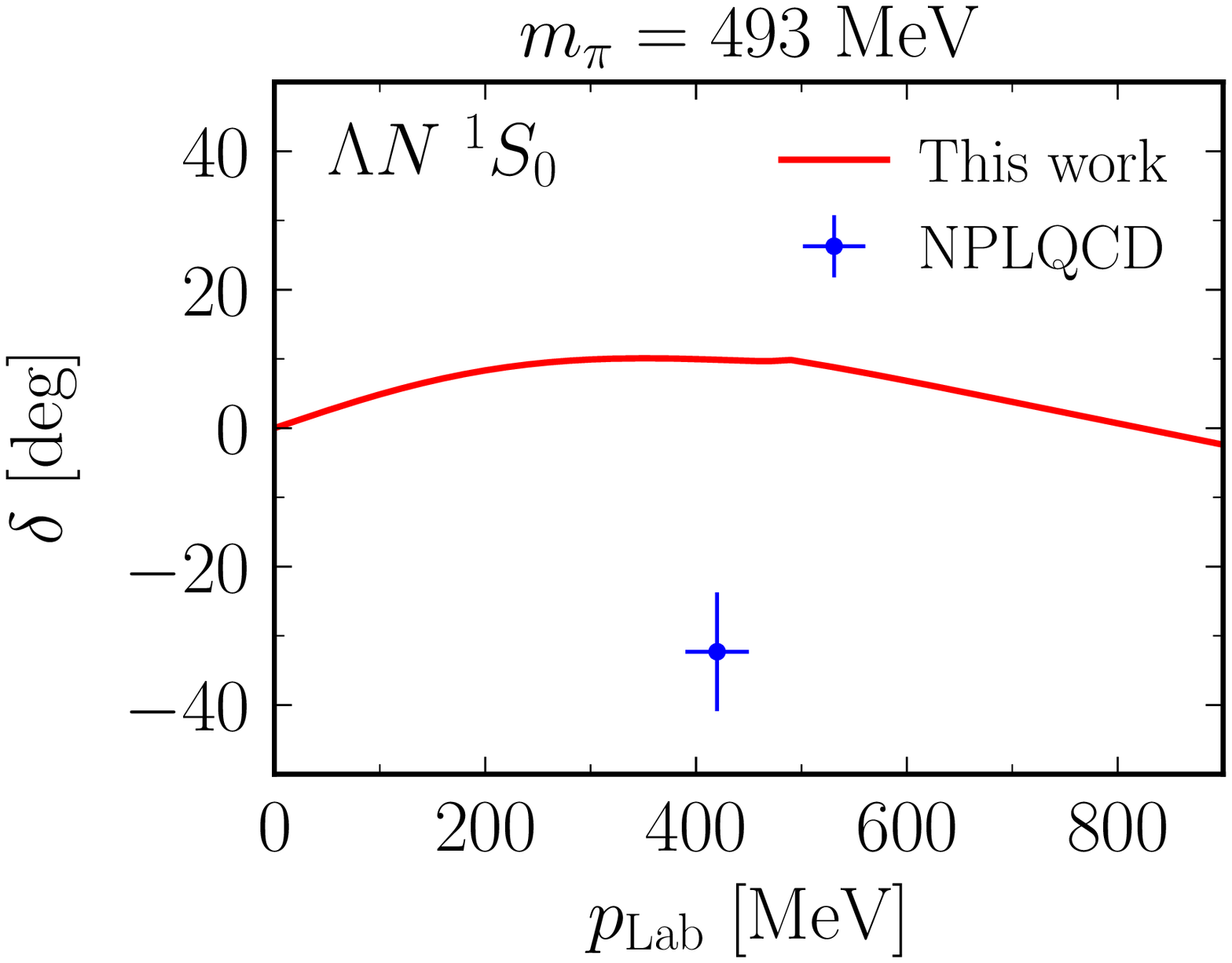}~~ \includegraphics[width=0.32\textwidth]{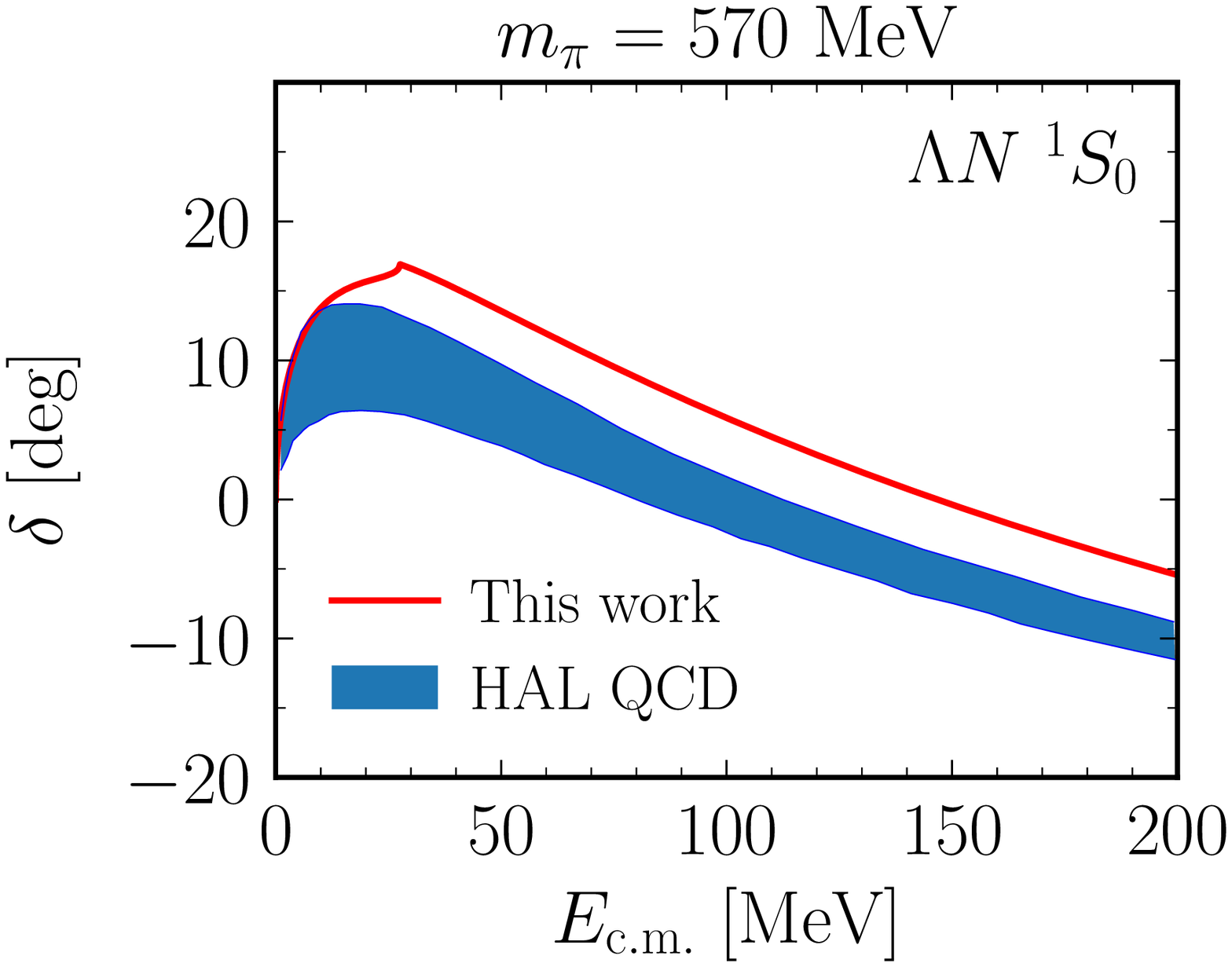}\\[0.2em]
\includegraphics[width=0.32\textwidth]{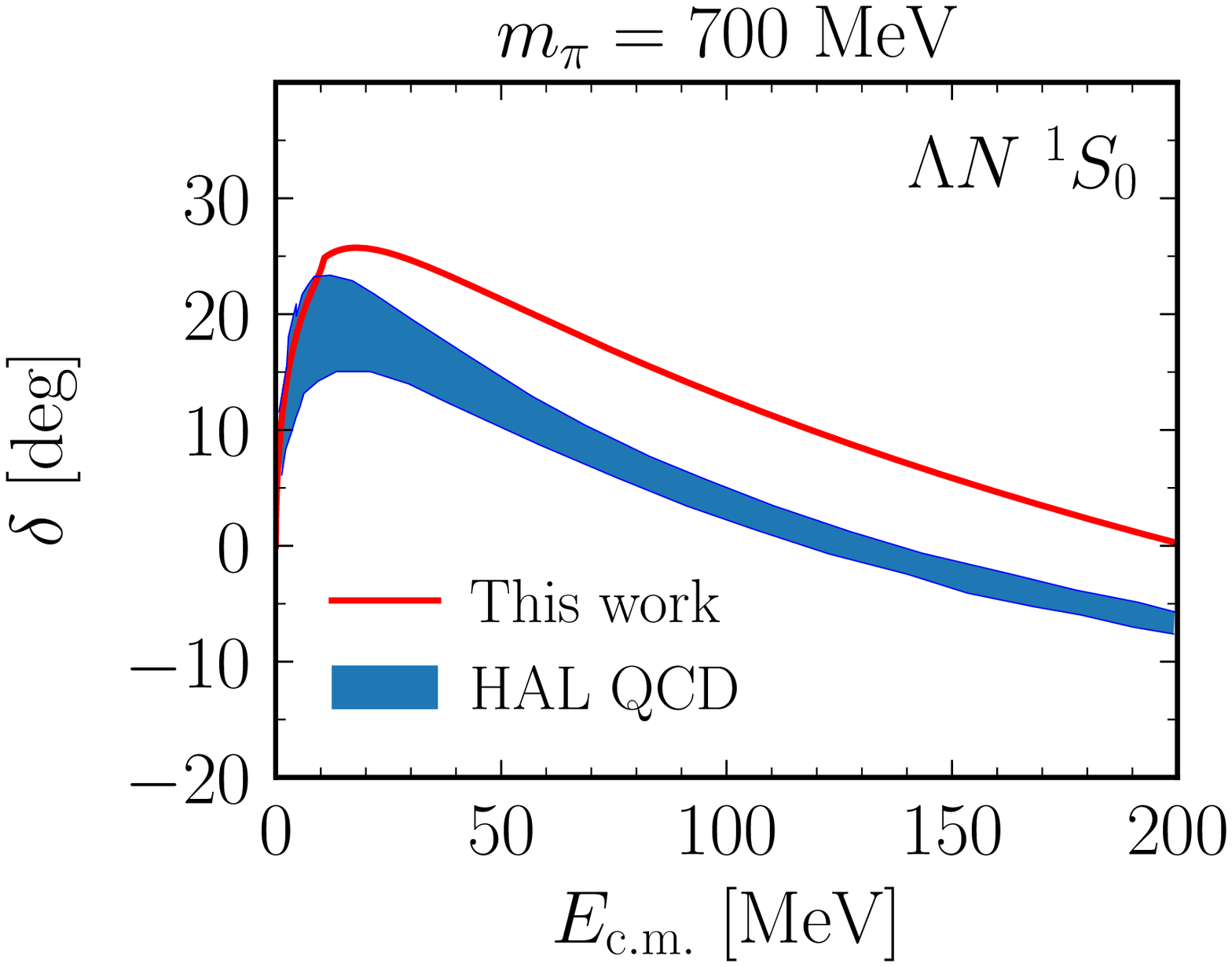}~~ \includegraphics[width=0.32\textwidth]{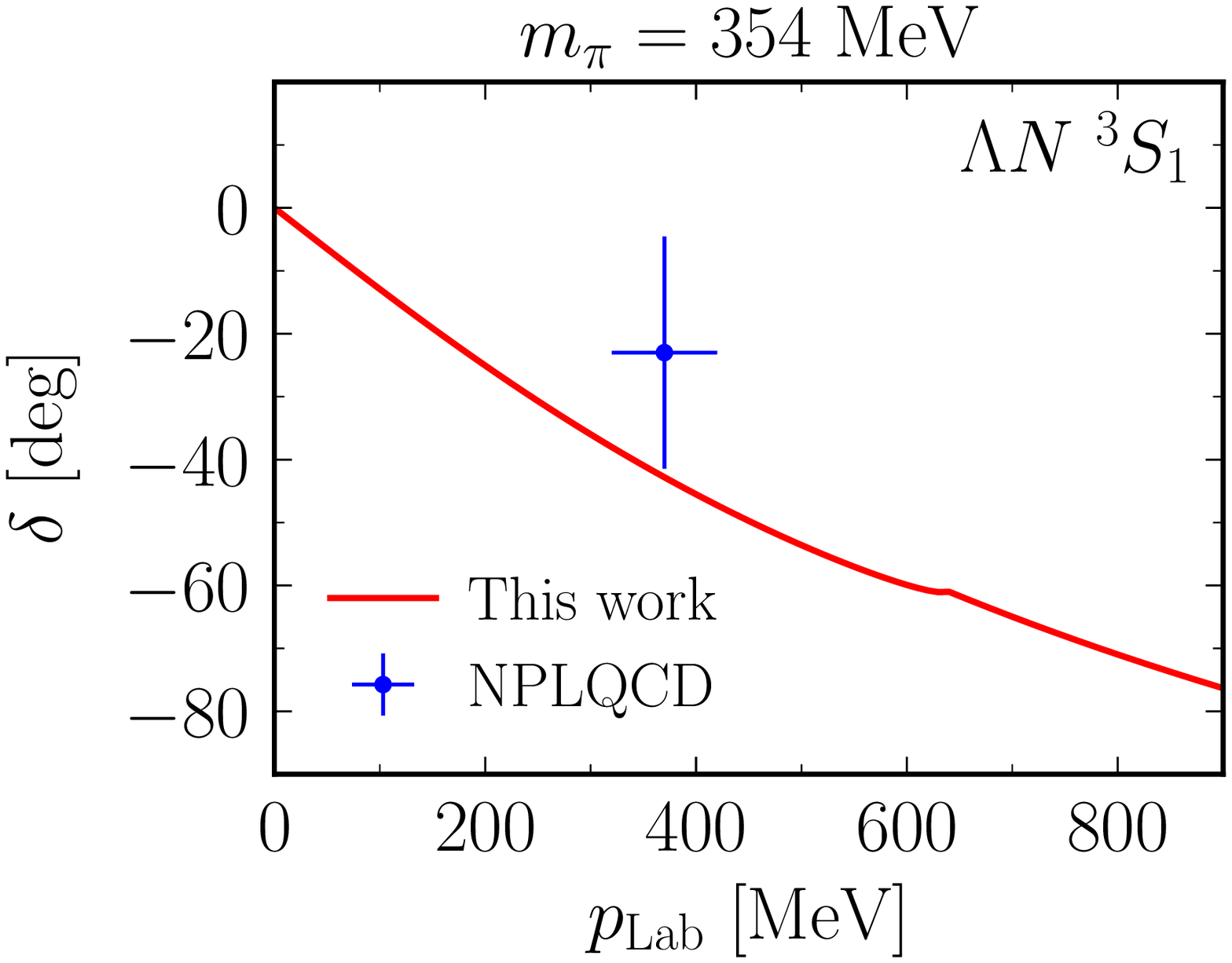} 
  \caption{$^1$S$_0$ and $^3$S$_1$ phase shifts of the $\Lambda N$ scattering for various values of the pion mass.}
  \label{Lattice_Extr}
\end{figure}

\medskip
Since we employ here an explicitly renormalizable formalism, there is
no implicit  quark-mass dependence of the renormalized LECs, see
Ref.~\cite{Epelbaum:2013ij} for more details. This allows us to
calculate the phase shifts of various processes for unphysical values
of the  
quark masses and to confront them with the results of lattice QCD
calculations.  
The $^1$S$_0$ and $^3$S$_1$ phase shifts of the $\Lambda N$ scattering for various values of the pion mass are shown in Fig.~\ref{Lattice_Extr} together with the results of the NPLQCD \cite{Beane:2006gf} and HAL QCD 
collaborations \cite{Miyamoto:2016hqo}.
{Note that the values of masses of pseudoscalar mesons and octet baryons are taken from the LHPC \cite{WalkerLoud:2008bp} and PACS-CS \cite{Aoki:2008sm} collaborations, which are tabulated in Table~\ref{Tab:LMass}, because the NPLQCD and HAL QCD employ the lattice configuartions of LHPC and PACS-CS, respectively.}
For $^1$S$_0$ partial wave, our results for large values of the pion
mass are in good agreement with those found by the HAL QCD
collaboration,\footnote{Notice, however, that the
  applicability of chiral EFT for such large values of the pion mass
  is rather questionable.} and the ones for $^3$S$_1$ partial wave
agree, within 
errors, with the prediction of the NPLQCD collaboration. 
\begin{table}[bt]
\caption{Masses of the pseudoscalar mesons and octet baryons obtained by the LHPC~\cite{WalkerLoud:2008bp} and PACS-CS~\cite{Aoki:2008sm} collaborations. 
   The first error is the statistical uncertainty and the second is determined by the lattice spacing.}
   \label{Tab:LMass}
\centering
\begin{tabular*}{\textwidth}{@{\extracolsep{\fill}}cccccc}
\hline
\hline
$M_\pi$(MeV) & $M_K$(MeV) & $m_N$(MeV) & $m_\Lambda$(MeV) & $m_\Sigma$(MeV) \\
 \hline
 LHPC & & & &  \\
355.9 & 602.9 & 1157.8(6.4)(23.1) & 1280.2(4.8)(25.6) & 1350.2(4.8)(27.0)\\ %& 1432.9(3.2)(28.6)\\
495.1 & 645.2 & 1288.2(6.4)(25.8) & 1369.3(4.8)(27.4) & 1409.1(6.4)(28.2)\\ %& 1469.5(4.8)(29.4)\\
596.7 & 685.6 & 1394.8(6.4)(27.9) & 1440.9(8.0)(28.8) & 1463.1(9.5)(29.2) \\ %& 1504.5(8.0)(30.1)\\
\hline
PACS-CS & & & & \\
569.7 & 713.2 & 1411.1(12.2)(21.8) & 1503.8(9.8)(23.2) & 1531.2(11.1)(23.6) \\ %& 1609.5(9.4)(24.8)\\
701.4 &789.0 & 1583.0(4.8)(24.4) & 1643.9(5.0)(25.4) & 1654.5(4.4)(25.5) \\ %& 1709.6(5.4)(26.4)\\
\hline
\hline
\end{tabular*}
\end{table}

\section{ Nonperturbative inclusion of higher-order contact
  interaction potentials in P-waves}
\label{rensubtract}

As already pointed out above, our results indicate that at least some
parts of subleading contrubutions to the potential will have to be
treated nonperturbatively. This then raises the question of whether
the scattering amplitude can still be renormalized, and the
infinite-cutoff limit can be taken. Fortunately, at least for the case
of nonperturbatively treated short-range interactions, the
renormalization program can still be carried out as will be
demonstrated below.  
We start by considering the case of the   $^3$P$_0$ partial waves and 
%To improve the description of the phase shifts by subtracting finite large contributions from the iterations of the LO long range potential, we  
%add the lowest-order contact interaction terms to the one pseudoscalar-meson-exchange coupled channel potential. Thus we 
treat non-perturbatively the following potential
\bea
\label{LO}
V \left(p_1',p_2'; p_1,p_2 \, \right)  =  
V_C(p_1',p_2';p_1,p_2) + V_{{\rm LO},M_P}^{IJ,KL} ,
\eea
where $V_{{\rm LO},M_P}^{IJ,KL}$ stands for the OME  projected onto the $^3P_0$ partial wave and the contact interaction potential is given by
\begin{equation}
V_C= \xi(p_1',p_2') \,{\cal C} \xi(p_1,p_2) , \ \ \ {\cal C} =  \left(%
\begin{array}{cc}
C_{\Lambda N,\,\Lambda N} & C_{\Lambda N,\,\Sigma N}  \\
C_{\Lambda N,\, \Sigma N}  & C_{\Sigma N,\,\Sigma N}  \\
\end{array}%
\right), \ \ \
\xi(p_1,p_2) = 
\left(%
\begin{array}{cc}
p_1 & 0  \\
0  & p_2  \\
\end{array}%
\right) .
\label{cxi}
\end{equation}
Using the results of Ref.~\cite{Kaplan:1996xu}
we write the solution to
the system of  integral equations for the potential of Eq.~(\ref{LO}) in the form,
which allows one to carry out the subtractive renormalization.
We start by writing the integral equations \eqref{PWEHDR} symbolically as 
\begin{equation}
T=V+V\,G\,T,
\label{eqsim}
\end{equation}
and rewrite it in the form \cite{Epelbaum:2015sha}
\begin{equation}
T=T_M+(1+T_M\,G)\,T_C (1+G\,T_M),
\label{OEQ12}
\end{equation}
where the amplitude $T_M$ satisfies the equation
\begin{equation}
T_M =V_{{\rm LO},M_P} + V_{{\rm LO},M_P} \,G\,T_M \,.
\label{OEQ1}
\end{equation}
For the contact interaction potential of Eq.~(\ref{cxi})
%\begin{equation}
%V_C(p'_1,p'_2; p_1,p_2)= \xi(p'_1,p'_2)\, {\cal C}\, \xi(p_1,p_2),
%\label{nuCfact}
%\end{equation}
the amplitude $T_C$ is given as
\begin{equation}
T_C(p'_1,p'_2;p_1,p_2)= \xi(p'_1,p'_2) {\cal X}\xi(p_1,p_2) ,
\label{chiCfact}
\end{equation}
where 
\begin{equation}
{\cal X}= \left[{\cal C}^{-1}-\xi\,G\,\xi -  \xi\,G\,T_M G\,\xi \,\right]^{-1}.
\label{chisol}
\end{equation}
Thus, the final expression for the amplitude $T$ has the form
\begin{equation}
T=T_M+(1+T_M\,G)\,\xi\,{\cal X}\,\xi (1+G\,T_M).
\label{taup}
\end{equation}
Analogously  to Refs.~\cite{Epelbaum:2015sha} and \cite{Baru:2019ndr} we apply the
subtractive (BPHZ-type) renormalization, i.e. we subtract {\it all} divergences in {\it all} loop diagrams 
 and regard the coupling constants as renormalized
 (i.e. cutoff-independent but renormalization scheme dependent) finite
 quantities, see e.g.~Ref.~\cite{Collins:1984xc} for  details
 of the BPHZ renormalization.   
Subtractive renormalization of the amplitude in Eq.~(\ref{taup})
corresponds to the inclusion of contributions of an infinite number of
counter terms generated by an infinite number of bare parameters of
the  effective Lagrangian \cite{Epelbaum:2018zli}. 

To apply subtractive renormalization to Eq.~(\ref{taup})
we notice that the amplitude $T_M$ and expressions $\bar\Xi(p'_1,p'_2)=(1+T_M\,G)\,\xi$  and $\Xi(p_1,p_2)=\xi (1+G\,T_M)$ are finite. 
Therefore we need to apply subtractions only to the quantity ${\cal X}$.  
While the $\xi\,G\,T_M G\, \xi$-term in Eq.~(\ref{chisol})  contains
only the overall logarithmic divergences,  the term  $\xi\,G\, \xi $
is quadratically divergent and, therefore, requires two additional BPHZ
subtractions.   

After carrying out the subtractions, the final  expression of the amplitude reads
\begin{equation}
T(p'_1,p'_2;p_1,p_2)=T_M(p'_1,p'_2;p_1,p_2)+ \bar \Xi(p'_1,p'_2)\, \frac{1}{ {\cal C}_R^{-1} - (\xi\,G\,\xi)^R - (\xi\,G\,T_M G\,\xi-\alpha)} \, \Xi(p_1,p_2) \,,
\label{tRen}
\end{equation}
where the counter term matrix $\alpha$ subtracts the overall divergences of $\xi\,G\,T_M G\,\xi$. The subtracted expression of $(\xi\,G\,\xi)^R$ is given by
\begin{equation}
(\xi\,G\,\xi)^R  =  \left(%
\begin{array}{cc}
I_{\Lambda N}\,, & 0 \\
0\,, & I_{\Sigma N}  \\
\end{array}%
\right),
\label{GchiiR}
\end{equation}
where the integrals $I_{IJ}$, subtracted at $E=E_0 < m_I+m_J$, are given by
\begin{eqnarray}
I_{IJ} &=&  \int \frac{ dk\, k^2}{2 \pi^2} \frac{k^2 (E_0-E)^3 m_I m_J}{\sqrt{m_i^2+k^2}
   \sqrt{m_J^2+k^2}
   \left(E_0-\sqrt{m_I^2+k^2}-\sqrt{m_J^2+k^2}\right)^3}  \nonumber\\
   &\times & \frac{1}{ \left(E-\sqrt{m_I^2+k^2}-\sqrt{m_J^2+k^2}-i \epsilon
   \right)} .
\label{IntegralS}
\end{eqnarray}
The imaginary part of this integral has the form
\begin{eqnarray}
\mathrm{Im}(I_{IJ}) &=& -\frac{m_I m_J \left[ m_I^4 -2 m_I^2
   \left(E^2+m_J^2\right)+\left(E^2-m_J^2\right){}^2
   \right]{}^{3/2}}{16 \pi  E^4}\,.
\label{ImPart}
\end{eqnarray}

\begin{figure}[t]
\includegraphics[width=0.245\textwidth]{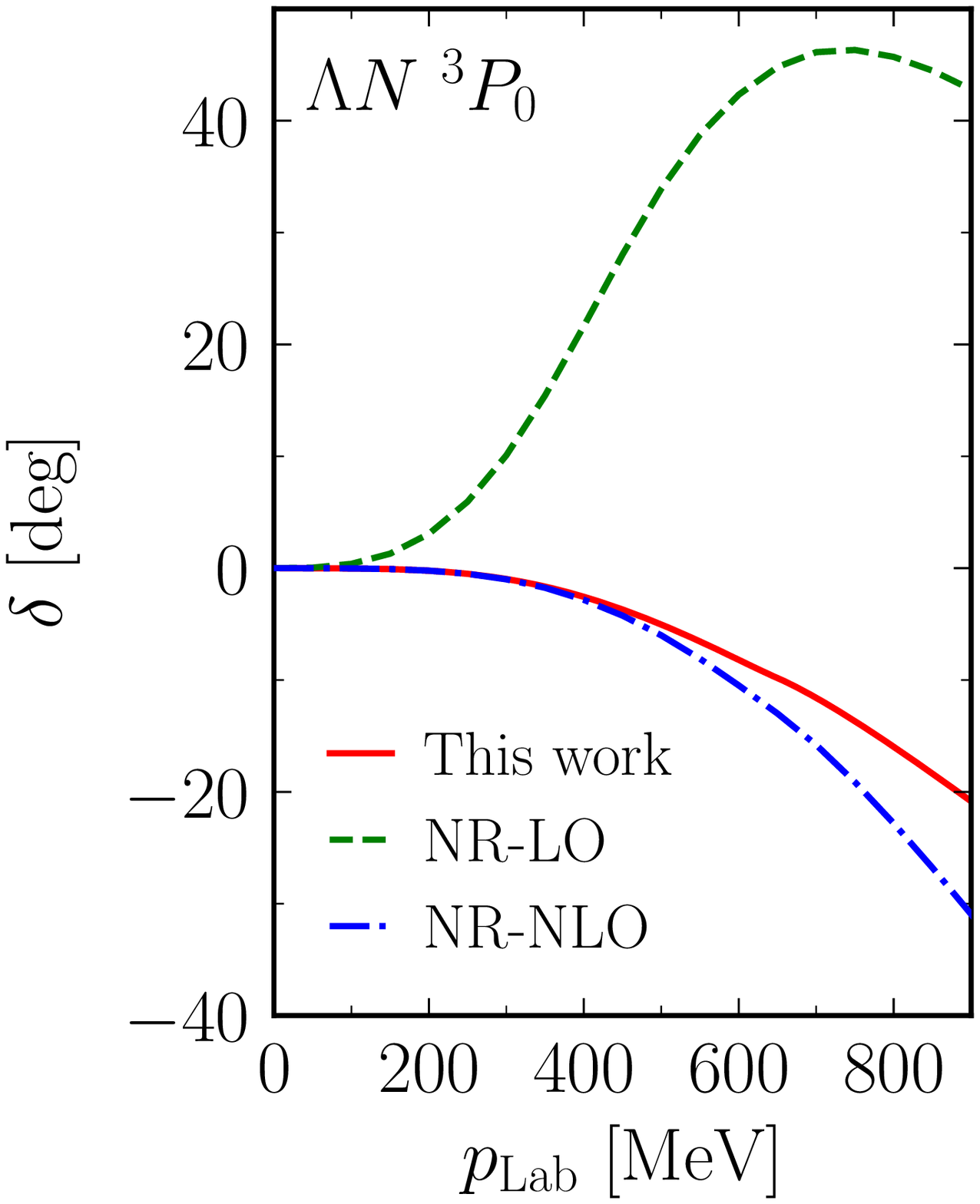}~~~~~
\includegraphics[width=0.25\textwidth]{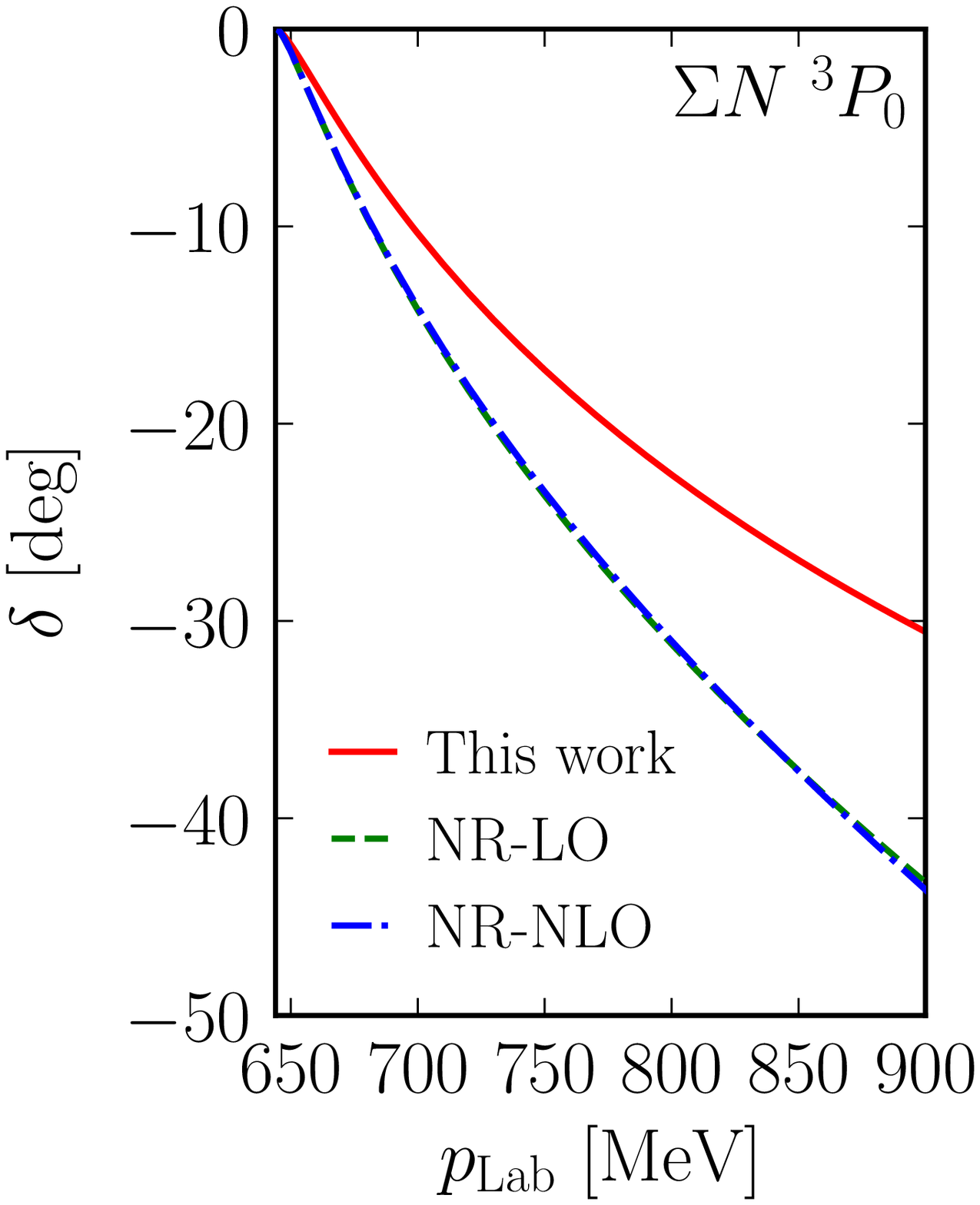}
  \caption{$^3$P$_0$ phase shifts of $\Lambda N$ and $\Sigma N$ scatterings with the  subtractive renormalization.}
  \label{3P0ren:fig}
\end{figure}

In practice, we fix the bare constant matrix $1/C=1/C_R+\alpha$ as a function of  the
cutoff numerically in such a way that it cancels the divergent part of
$\xi\,G\,T_M G\,\xi$ and the resulting cutoff-independent coupled-channel 
scattering amplitudes describe the phase shifts for a fixed value of
the energy.
The three renormalized LECs are fixed by the low-energy $^3$P$_0$
phase shifts of the $\Lambda N$ and $\Sigma N$ scatterings and the inelasticity parameters of Ref.~\cite{Haidenbauer:2019boi}.
The results
are shown in Fig.~\ref{3P0ren:fig}. The description of the $\Lambda N$
$^3$P$_0$ phase shifts is satisfactory while the results of $\Sigma N$
$^3$P$_0$ phase shifts are similar to the ones of the OME potential.

\begin{figure}[tb]
\includegraphics[width=0.25\textwidth]{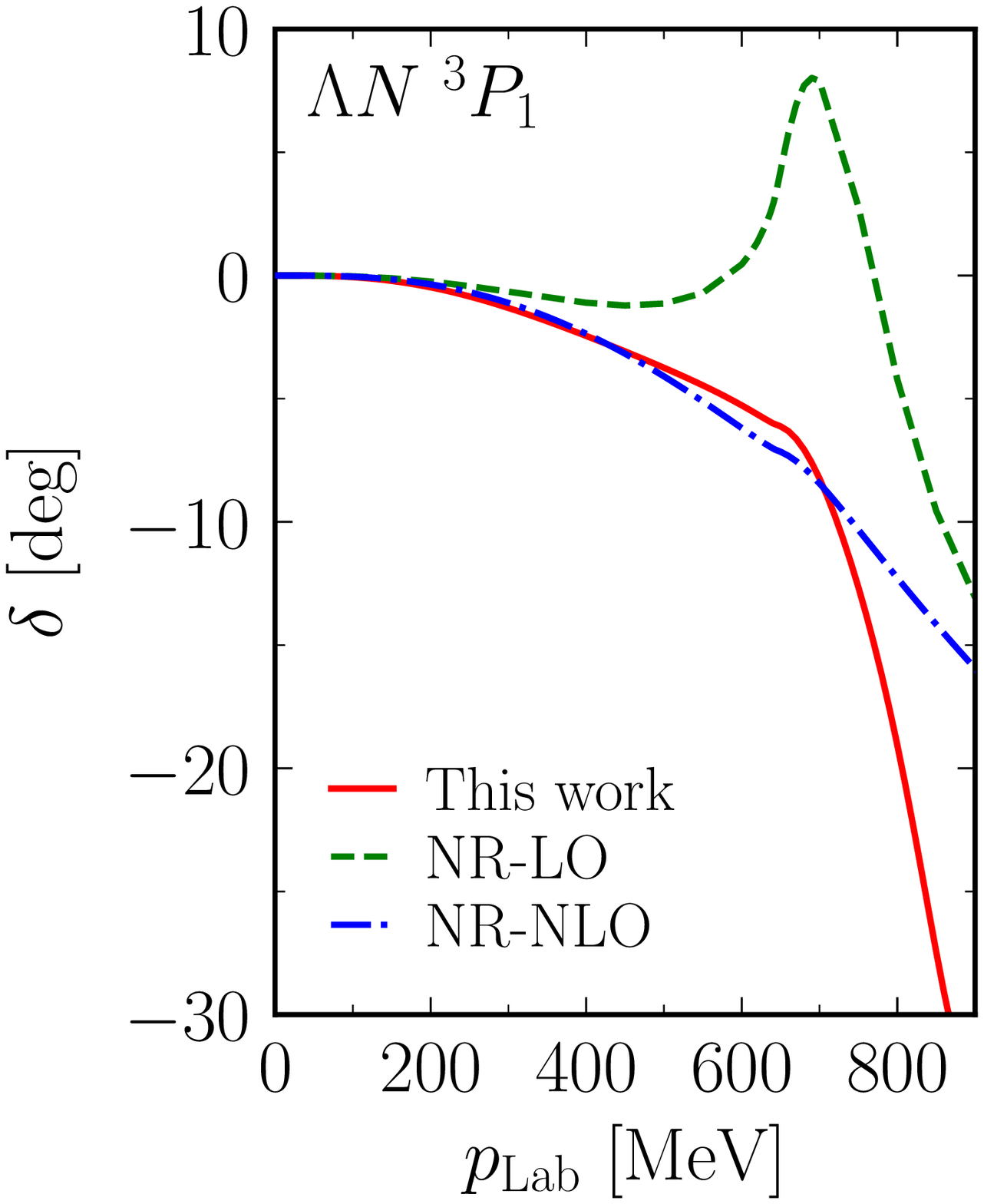}~
\includegraphics[width=0.25\textwidth]{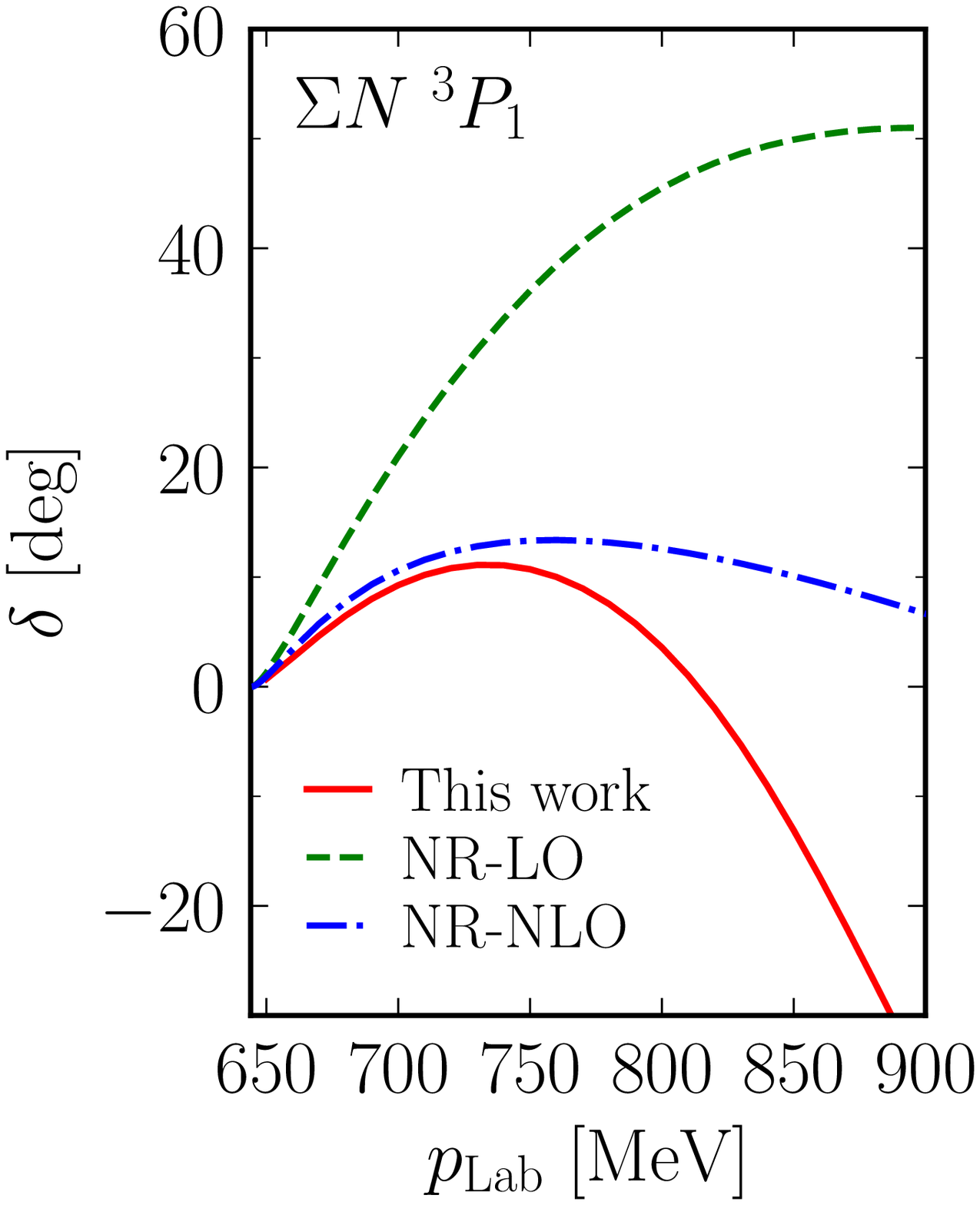}~ \includegraphics[width=0.25\textwidth]{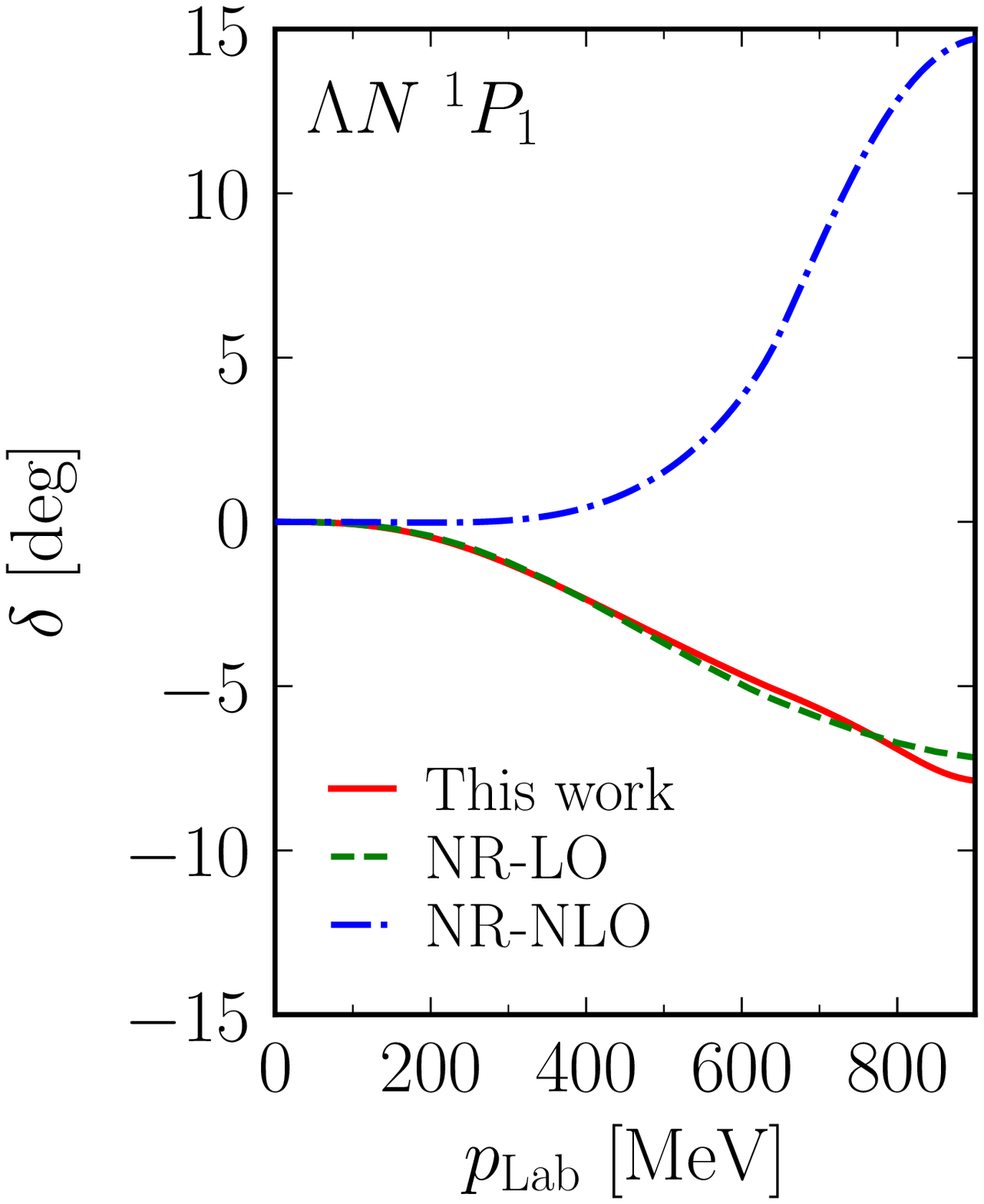}~
\includegraphics[width=0.25\textwidth]{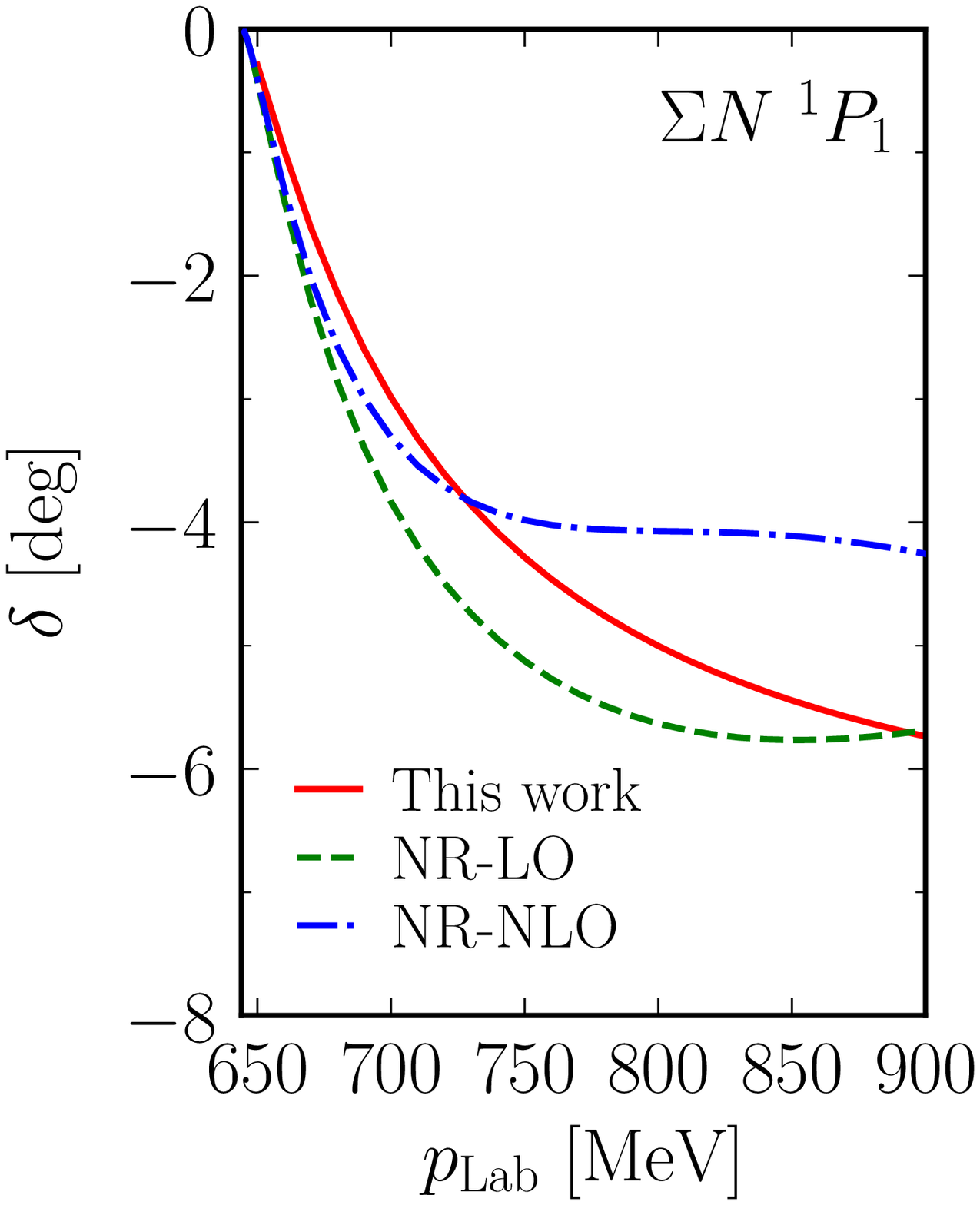}
  \caption{$^3$P$_1$ phase shifts of $\Lambda N$ and $\Sigma N$ scatterings with the subtractive renormalization.}
  \label{3P1-ren:Fig}
\end{figure}

We also extended the renormalization procedure to the $^1$P$_1$-$^3$P$_1$ coupled channels. 
We treat nonperturbatively three contact interactions for the coupled
 particle channels in $^3$P$_1$ partial wave and fix the corresponding
renormalized LECs via the description of the low-energy $^3$P$_1$
phase shifts of Ref.~\cite{Haidenbauer:2019boi}.  
As seen from the results in Fig.~\ref{3P1-ren:Fig}, a rather good
description of $^3$P$_1$ partial wave phase shifts is achieved. One
should also notice that there is small effect on the $^1$P$_1$ partial
wave phase shifts.

\section{Summary}
\label{conclusions}

In this paper we calculated the lambda-nucleon scattering amplitude of the strangeness $S=-1$ sector in the framework
of manifestly Lorentz-invariant formulation of SU(3) BChPT  by 
applying time-ordered perturbation theory \cite{Baru:2019ndr}. 

 For the case of
baryon-baryon scattering, the relative importance of time-ordered diagrams can be determined using the
Weinberg's power counting rules \cite{Weinberg:rz,Weinberg:um}.
To sum up the relevant
contributions it is convenient to define the effective potential
as a sum of all two-baryon irreducible contributions to the scattering
amplitude within TOPT.
The scattering amplitudes are obtained as solutions of a system of the
coupled-channel
integral equations with the potentials at the corresponding order. 
These equations represent a coupled-channel generalization of the
Kadyshevsky equation 
\cite{kadyshevsky} and feature a milder ultraviolet behaviour as compared to
their non-relativistic analogs.  
By solving the integral equations for the LO amplitudes and including
corrections perturbatively one can remove cutoff-dependence from the
physical amplitudes.  
Also for higher-order contact interactions included nonperturbatively one can
remove all divergences by performing BPHZ type subtractions. This
corresponds to taking into account an infinite number of counter terms
of higher orders. 

The large discrepancy between the results of our LO calculations
and the  phase shifts from Ref.~\cite{Haidenbauer:2019boi} suggest that certain contributions to the BB potential beyond
LO must be treated nonperturbatively in the $^3P_0$ and the  $^3P_1$
partial waves. Thus, we have extended our calculations to include the NLO
short-range  interactions in these partial waves and carried out
subtractive  renormalization in a way consistent with
EFT.  The resulting phase shifts are found to be in a good agreement with the
corresponding  ones from the non-relativistic approach of  Ref.~\cite{Haidenbauer:2019boi}. We also
studied the quark mass dependence of the $^1S_0$ and $^3S_1$ phase
shifts and compared the resulting LO predictions with the available
results from the lattice QCD simulations.

\section*{Acknowledgments}
  We thank Johann Haidenbauer for providing the non-relativistic
  chiral potentials from
  Refs.~\cite{Polinder:2006zh,Haidenbauer:2013oca,Haidenbauer:2019boi} and
  for numerous fruitful discussions. We are also grateful to
  Ulf-G. Mei{\ss}ner and  Johann Haidenbauer for their insightful and
  helpful comments on the manuscript.
This work was supported in part by the BMBF  (Grant
No.~05P18PCFP1), by the DFG and NSFC
 through funds provided to the
Sino-German CRC 110 ``Symmetries and the Emergence of Structure in QCD" (NSFC
Grant No.~11621131001, DFG Grant No.~TRR110) and by
the Georgian Shota Rustaveli National
Science Foundation (Grant No. FR17-354).% and the Russian Science Foundation (Grant No.~18-12-00226). 

\end{document}